\documentclass[dvipdfmx]{PoS}
\let\ifpdf\relax

\title{
  Partial restoration of chiral symmetry \\ inside hadrons
}  

\ShortTitle{
  Partial restoration of chiral symmetry inside hadrons
}

\author{\speaker{Takumi~Iritani}$^a$, Guido~Cossu$^b$, and Shoji~Hashimoto$^{bc}$ \\
  \llap{$^a$}Yukawa Institute for Theoretical Physics (YITP),
  Kyoto 606-8502, Japan \\
  \llap{$^b$}High Energy Accelerator Research Organization (KEK),
  Ibaraki 305-0801, Japan \\
  \llap{$^c$}School of High Energy Accelerator Science,
  The Graduate University for Advanced Studies (Sokendai),
  Ibaraki 305-0801, Japan \\
  E-mail: \email{iritani@yukawa.kyoto-u.ac.jp}
}

\abstract{
  By using the overlap-Dirac operator eigenmodes, we investigate
  spatial distribution of the chiral condensate around static color sources
  for both quark-antiquark and three quark systems.
  In the presence of color sources, a characteristic flux-tube structure 
  appears among them,
  suggesting a linear confining potential.
  We show that the magnitude of the condensate is reduced inside the color flux,
  which indicates the partial restoration of chiral symmetry
  inside the \textit{hadrons}.
  Considering a periodic box containing a static baryon source,
  which mimics the \textit{nuclear matter},
  we estimate the chiral symmetry restoration
  in the presence of finite baryon number density.
}

\FullConference{The 32nd International Symposium on Lattice Field Theory,\\
		23-28 June, 2014\\
		Columbia University New York, NY}

\begin{document}

\section{Introduction}

Chiral symmetry is spontaneously broken in the non-perturbative vacuum of QCD.
This symmetry breaking is closely related to underlying topological structure 
of the vacuum, such as the instantons.
One of the fundamental order parameters 
to characterize the vacuum is the chiral condensate
$\langle \bar{q}q \rangle$,
which has a non-zero expectation value in the vacuum.
How it is affected by finite temperature or/and density 
is an interesting question of the QCD dynamics.

Probing the vacuum with color charges
interesting non-trivial structures can be found.
For example,
a color flux-tube emerges between quark and antiquark,
which induces a linearly rising confining potential \cite{Bali:2000gf}.
This flux-tube can be observed by gluonic degrees of freedom
around the color sources from lattice QCD calculations
\cite{Bali:1995,Iritani:2013}.
It is also reported that
color charges modify
the chiral condensate \cite{Iritani:2013,Faber:1993,Suganuma:1990nn}.

In this work, 
we analyze the chiral condensation 
near static color sources
by using overlap-Dirac eigenfunctions.
We discuss the modification of the chiral condensate 
for both quark-antiquark and three-quark systems
made of static quark lines as toy models of hadrons.
Considering a single such toy baryon in a periodic box,
we can also model the nuclear matter, with which
we estimate the size of chiral symmetry restoration at finite density.

\section{Partial restoration of chiral symmetry inside hadrons}

In this study, we use 2+1-flavor dynamical overlap-fermion configurations,
which are generated by the JLQCD Collaboration \cite{JLQCD}.
The massless overlap-Dirac operator is defined by
\begin{equation}
  D_\mathrm{ov}(0) = m_0 \left[ 1 + \gamma_5 \mathrm{sgn} \ H_W(-m_0) \right],
  \label{}
\end{equation}
where $H_W(-m_0) = \gamma_5 D_W(-m_0)$ is the hermitian Wilson-Dirac operator,
and sgn denotes the matrix sign function \cite{GinspargWilson,Neuberger:1998}.
The overlap-fermion is suited for
the study of chiral symmetry
and topological properties of QCD \cite{JLQCD},
as it preserves exact symmetry on the lattice \cite{GinspargWilson}.
We use $16^3 \times 48$ and $24^3 \times 48$ lattices 
at $\beta = 2.3$, which corresponds to a lattice spacing $a^{-1} = 
1.759(10)$ GeV.  The dynamical quark masses are $m_\mathrm{ud} = 0.015a^{-1}$
and $m_\mathrm{s} = 0.080a^{-1}$, and the global topological charge 
is fixed at $Q = 0$.

\subsection{Local chiral condensate $\bar{q}q(x)$}

In order to study the chiral condensation inside hadrons,
we measure the expectation value of the scalar density operator
$\bar{q}q({x})$ around static color sources.
By using the overlap-Dirac eigenfunction $\psi_\lambda(x)$
and the corresponding eigenvalue $\lambda$,
which satisfies $D_{\mathrm{ov}}(0) \psi_\lambda = \lambda \psi_\lambda$,
the quark-loop due to the insertion of $\bar{q}q(x)$ is written as
\begin{equation}
  \bar{q}q(x) = - \sum_\lambda \frac{\psi_\lambda^\dagger(x)\psi_\lambda(x)}{m_q
    + (1 - \frac{m_q}{2m_0})\lambda},
  \label{eq:localChiralCondensate}
\end{equation}
for a quark mass $m_q$.
The chiral condensate $\langle \bar{q}q \rangle$ is given 
by the spatial average of $\bar{q}q(x)$,
which leads to the well-known Banks-Casher relation \cite{BanksCasher}.
Instead of summing over all eigenmodes,
we introduce its low-mode approximation by truncating the sum from the lowest eigenvalue
to an upper limit, that is set by the number of modes $N$.

\begin{figure}[h]
  \centering
  \includegraphics[width=0.49\textwidth,clip]{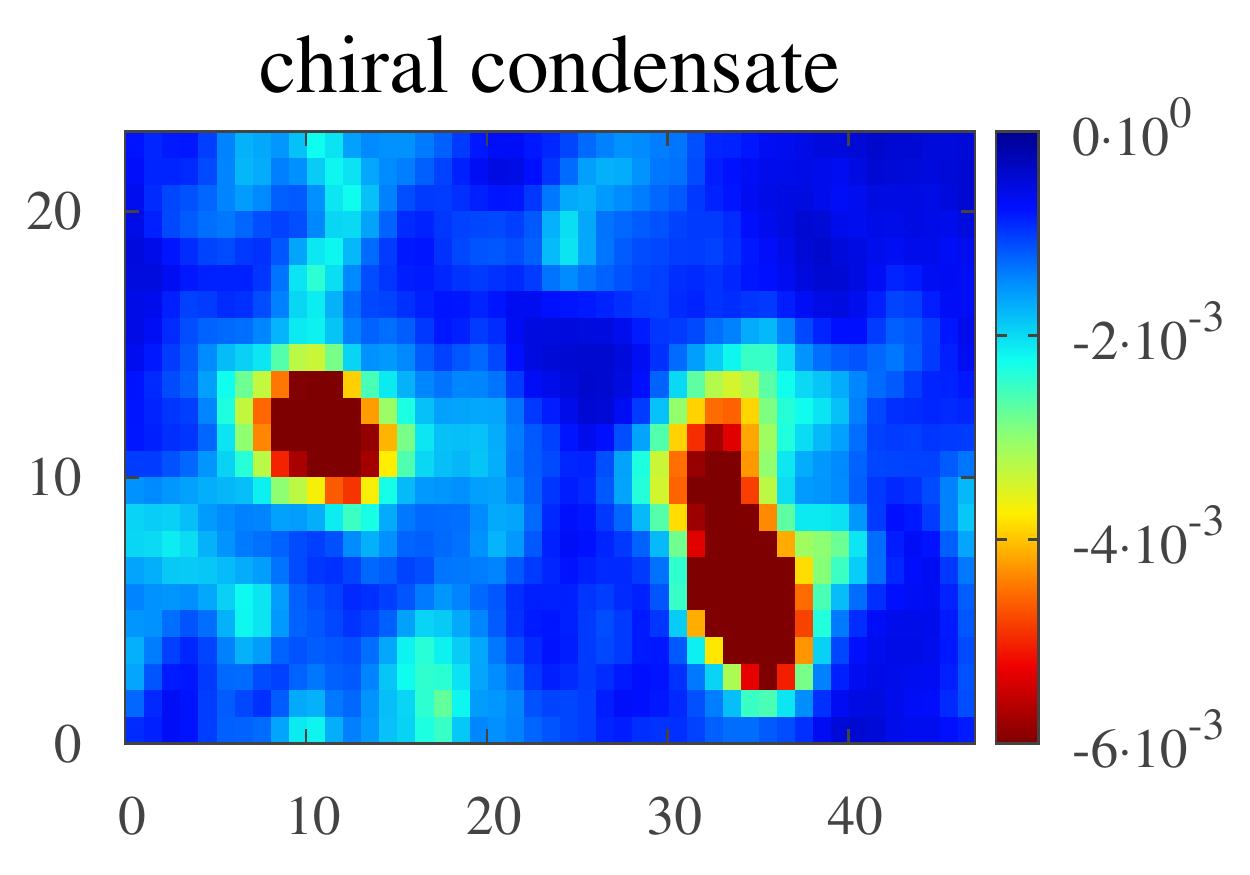}
  \includegraphics[width=0.49\textwidth,clip]{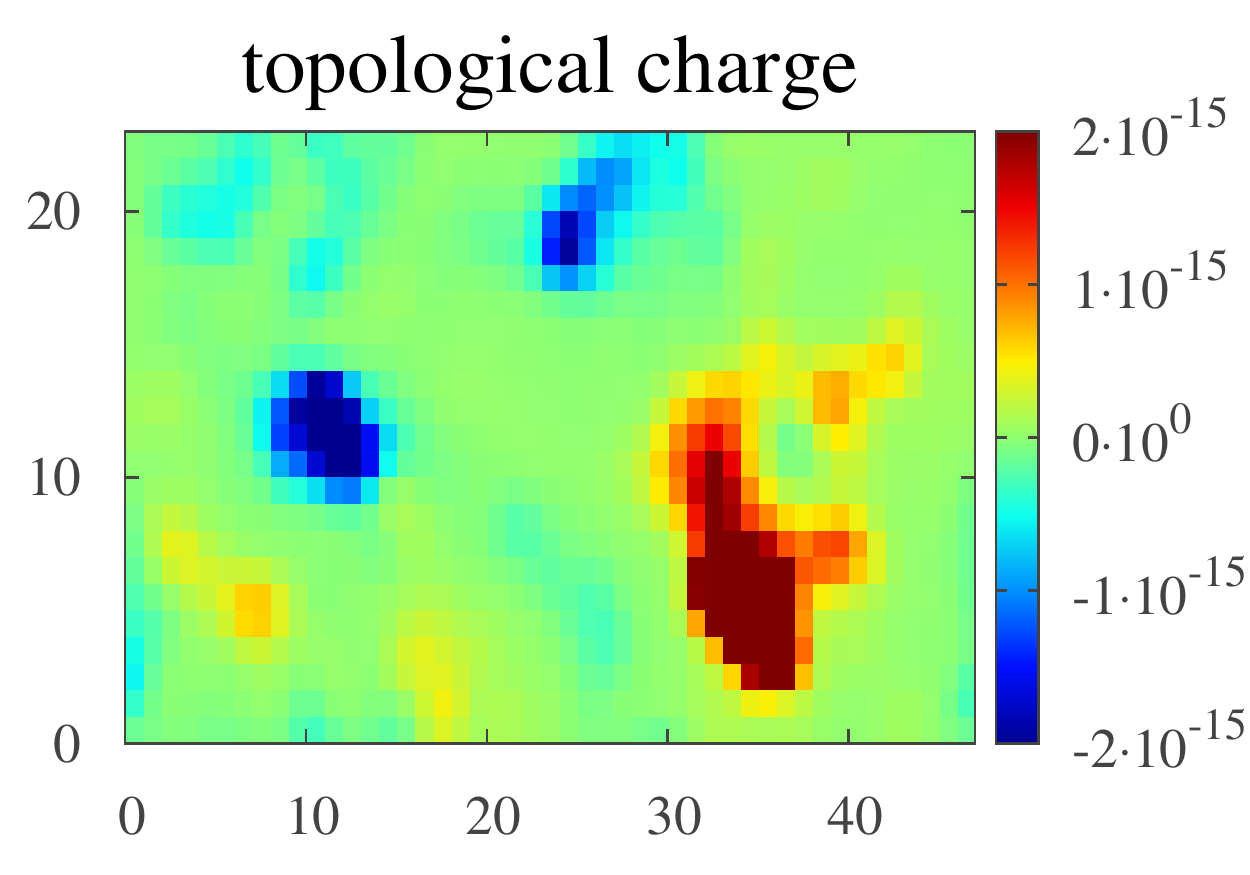}
  \caption{ \label{fig:local_chiral}
    A snapshot of the local chiral condensate and topological charge distribution
    as calculated using low-lying 20 eigenmodes.
    The topological charge density is constructed by
    the Dirac eigenfunction 
    using Gattringer's decomposition \cite{Iritani:2013,Gattringer:2002}.
    They show an intersection on the same $X$-$T$ slice of the four-dimensional
    lattice. Data for 2+1-flavor QCD at $\beta = 2.30$ on a $24^3 \times 48$ lattice.
  }
\end{figure}

We find that the local chiral condensate $\bar{q}q(x)$ forms clusters,
which seem to correlate with topological charge distribution (Fig.~\ref{fig:local_chiral}).
This result may suggest the instanton-based picture of QCD vacuum
\cite{Schafer:1996wv}.

\subsection{Chiral condensate in quark-antiquark system}

\begin{figure}
  \centering
  \includegraphics[width=0.93\textwidth,clip]{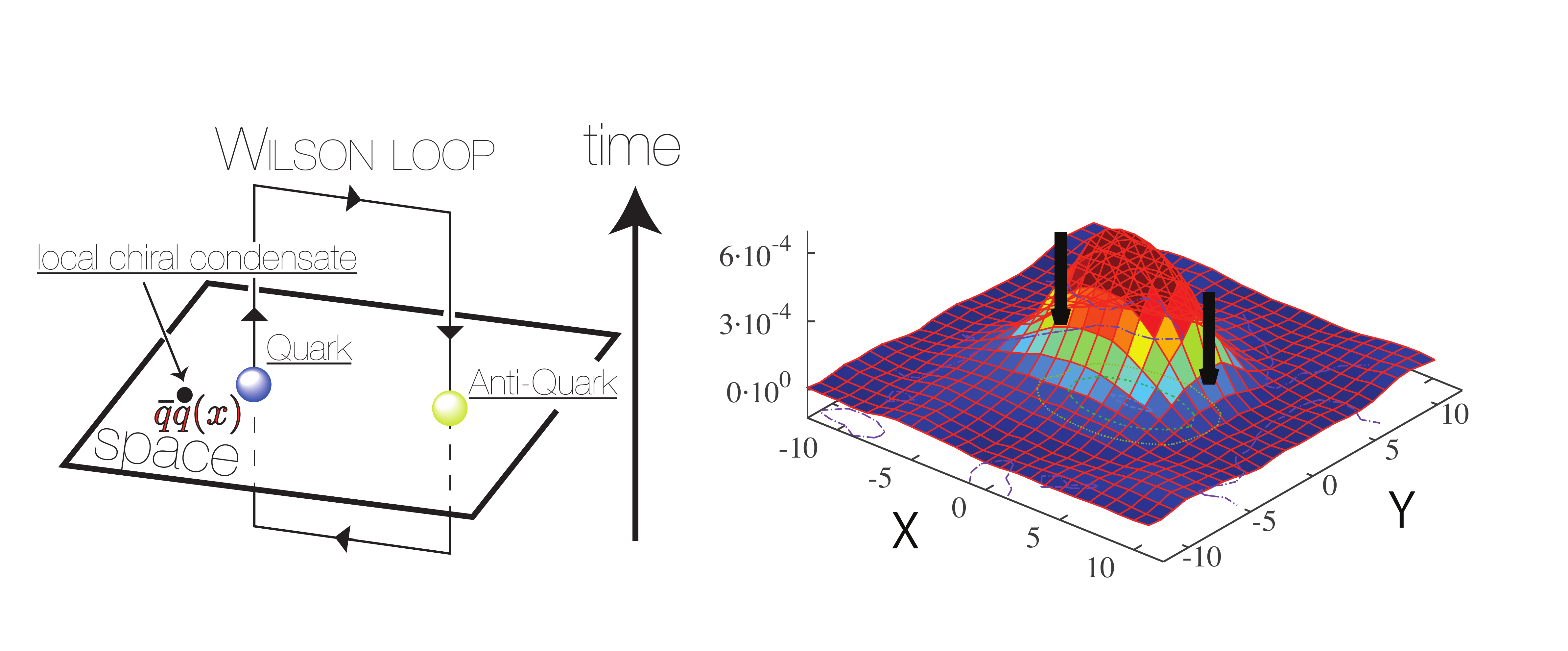}
  \caption{ \label{fig:chiral_around_qqbar}
  (a) A schematic picture of local chiral condensate measurement
  around quark-antiquark pair.
  (b) A change of local chiral condensate around quark-antiquark pair.
      The black bars denote the location of sources
      at $(X,Y) = (5,0)$ and $(-5,0)$.}
\end{figure}

Now, we discuss the modification of the chiral condensate around the static color sources.
We measure a spatial distribution of $\bar{q}q(\vec{x})$ around
the static color sources as
\begin{equation}
  \langle \bar{q}q(\vec{x}) \rangle_W 
  \equiv \frac{\langle\bar{q}q(\vec{x}) W(R,T)\rangle}{\langle W(R,T)\rangle}
    - \langle \bar{q}q \rangle,
  \label{eq:diffLocalChiral}
\end{equation}
where $W(R,T)$ is the Wilson loop with the size of $R\times T$, 
which corresponds to a static quark-antiquark pair with a separation $R$.
A schematic picture of the measurement is shown 
in Fig.~\ref{fig:chiral_around_qqbar} (a).

Figure~\ref{fig:chiral_around_qqbar} (b) is the spatial distribution of the local chiral 
condensate with $R = 10$.
The black bars show the positions of the color sources.
Here, we use a low-mode truncated condensate
$\bar{q}q^{(N)}(\vec{x})$ by truncating the sum at $N = 160$
in Eq.~(\ref{eq:localChiralCondensate}),
which is large enough to discuss the condensate \cite{Noaki:2009xi}.
The valence quark mass is $m_q = 0.015$.
In order to improve the signal of the Wilson loop, 
we adopt the APE smearing for the spatial link-variable,
and the temporal extension is set to $T = 4$ where the ground state becomes dominant.
The number of configurations is 50.

As shown in Fig.~\ref{fig:chiral_around_qqbar} (b),
there appears a tube-like structure between color sources,
where difference of condensate $\langle \bar{q}q(\vec{x}) \rangle_W$ becomes positive.
Since $\langle \bar{q}q \rangle$ is negative in the vacuum,
this result implies that the magnitude of chiral condensate is reduced inside
the $\mathrm{Q\bar{Q}}$-system.

In order to discuss the restoration quantitatively,
the renormalization of the operator $\bar{q}q$ is required.
By taking the mode truncation as a regularization scheme \cite{Noaki:2009xi},
the power divergence is parameterized as
\begin{equation}
  \langle \bar{q}q \rangle^{(N)} = 
  \langle \bar{q}q^{\rm (subt)} \rangle  
  + {c_1^{(N)}m_q/a^2} + c_2^{(N)}m_q^3,
  \label{eq:subtractChiral}
\end{equation}
in which the potential $1/a^3$ term is absent because of
the exact chiral symmetry of overlap-fermion \cite{GinspargWilson,Neuberger:1998}.
These coefficients $c_1^{(N)}$ and $c_2^{(N)}$ may be obtained 
by fitting $\langle \bar{q}q \rangle^{(N)}$
as a function of $m_q$.
The subtracted condensate $\langle \bar{q}q^{(\rm subt)} \rangle$ becomes finite up to logarithmic renormalization.
By taking the ratio,
\begin{equation}
  r(\vec{x}) \equiv \frac{\langle \bar{q}q^{\rm (subt)}(\vec{x})W(R,T) \rangle}{
    \langle \bar{q}q^{\rm (subt)} \rangle\langle W(R,T) \rangle},
  \label{eq:ratioCondensate}
\end{equation}
the remaining divergence is also canceled.

\begin{figure}
  \centering
  \includegraphics[width=0.47\textwidth,clip]{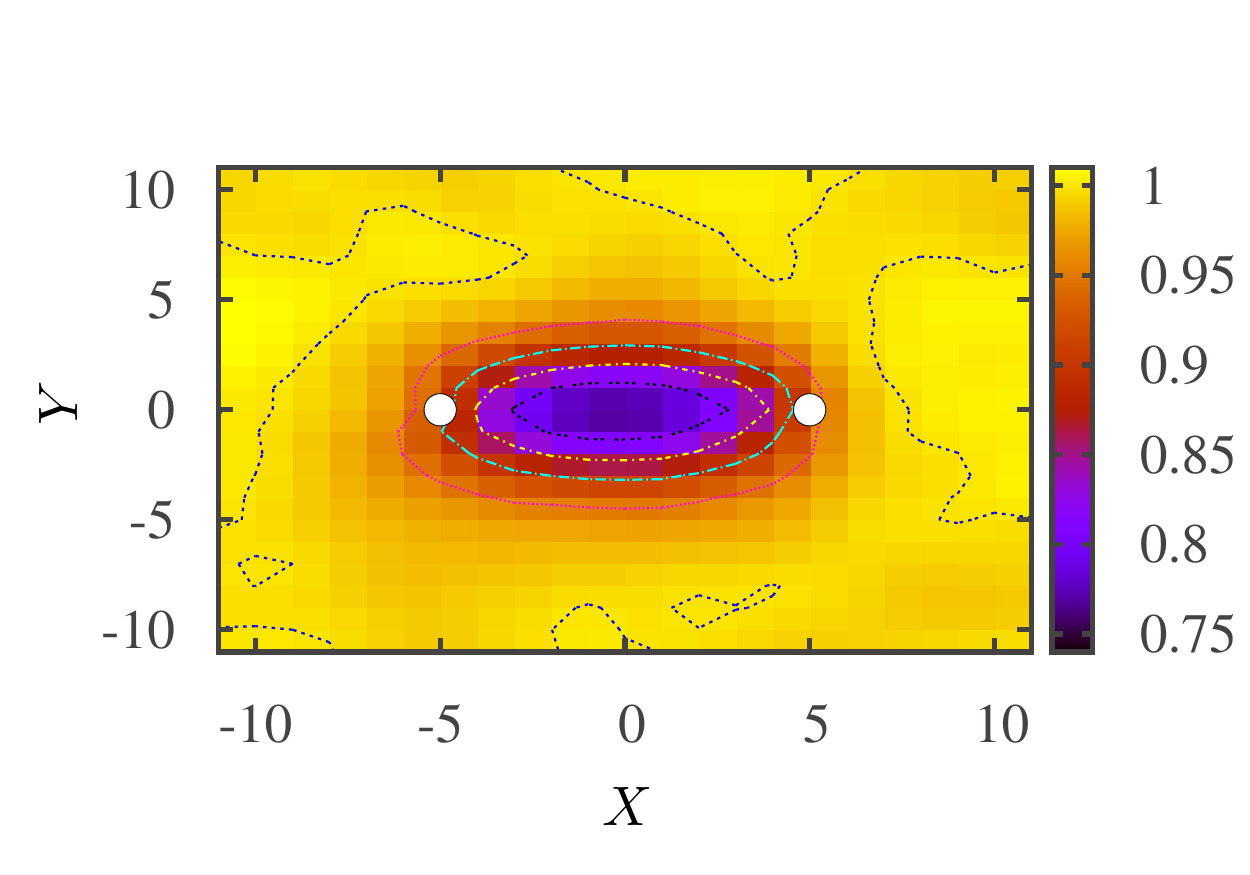}
  \includegraphics[width=0.47\textwidth,clip]{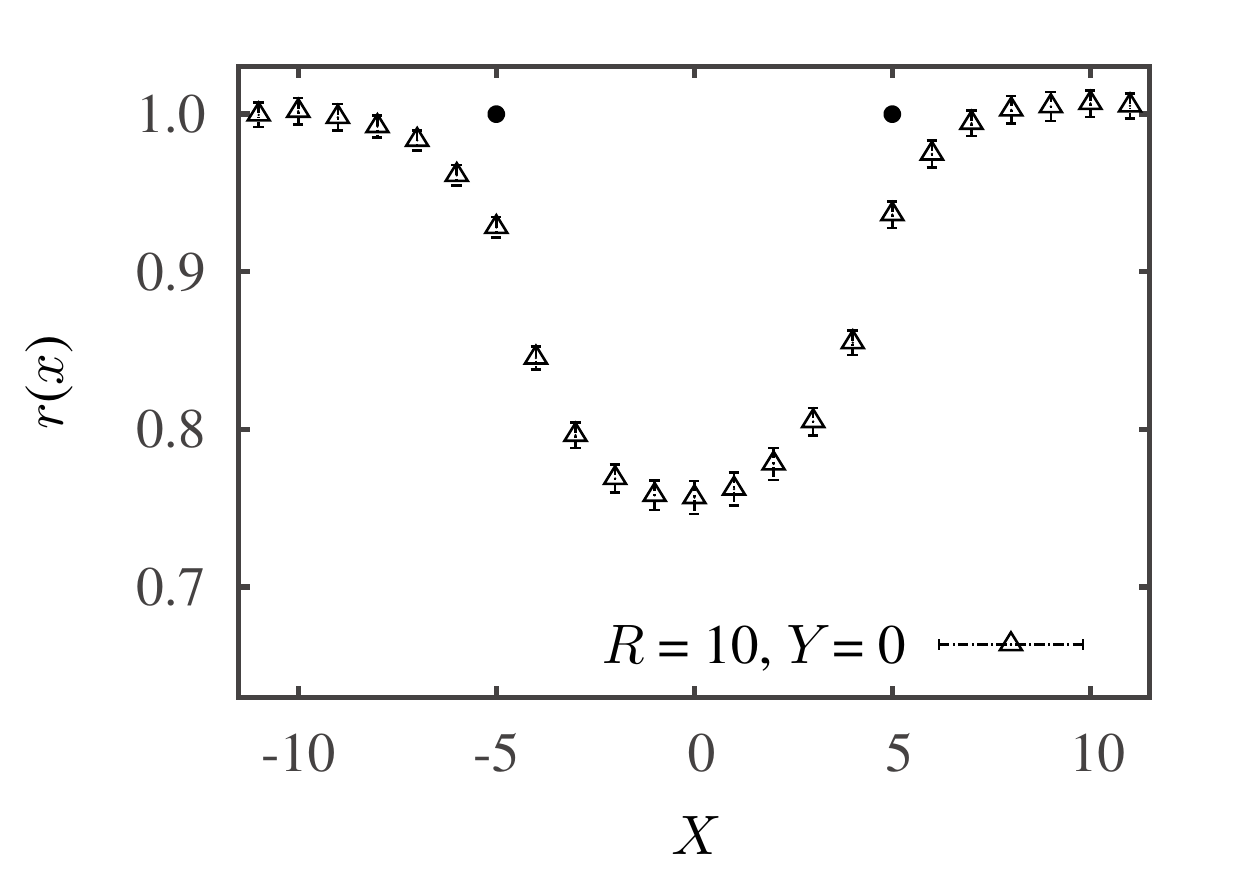}
  \caption{
    \label{fig:chiral_ratio}
    (a) The heat-map of chiral condensate ratio $r(\vec{x})$ around the static color sources,
    whose locations are depicted by circles.
    The magnitude of condensate becomes smaller between color sources.
    (b) The cross-section of $r(\vec{x})$ along static color sources,
    which are denoted by the black points.
  }
\end{figure}

Figure~\ref{fig:chiral_ratio} shows a heat-map of the ratio $r(\vec{x})$
and its cross-section along the flux with $R = 10$.
There appears a tube-like region, where magnitude of the chiral condensate is reduced.
In particular, 
the restoration becomes largest around the center of the color charges.
In this case, the condensate is reduced by about 25\% at the center.

\begin{figure}
  \centering
  \includegraphics[width=0.49\textwidth,clip]{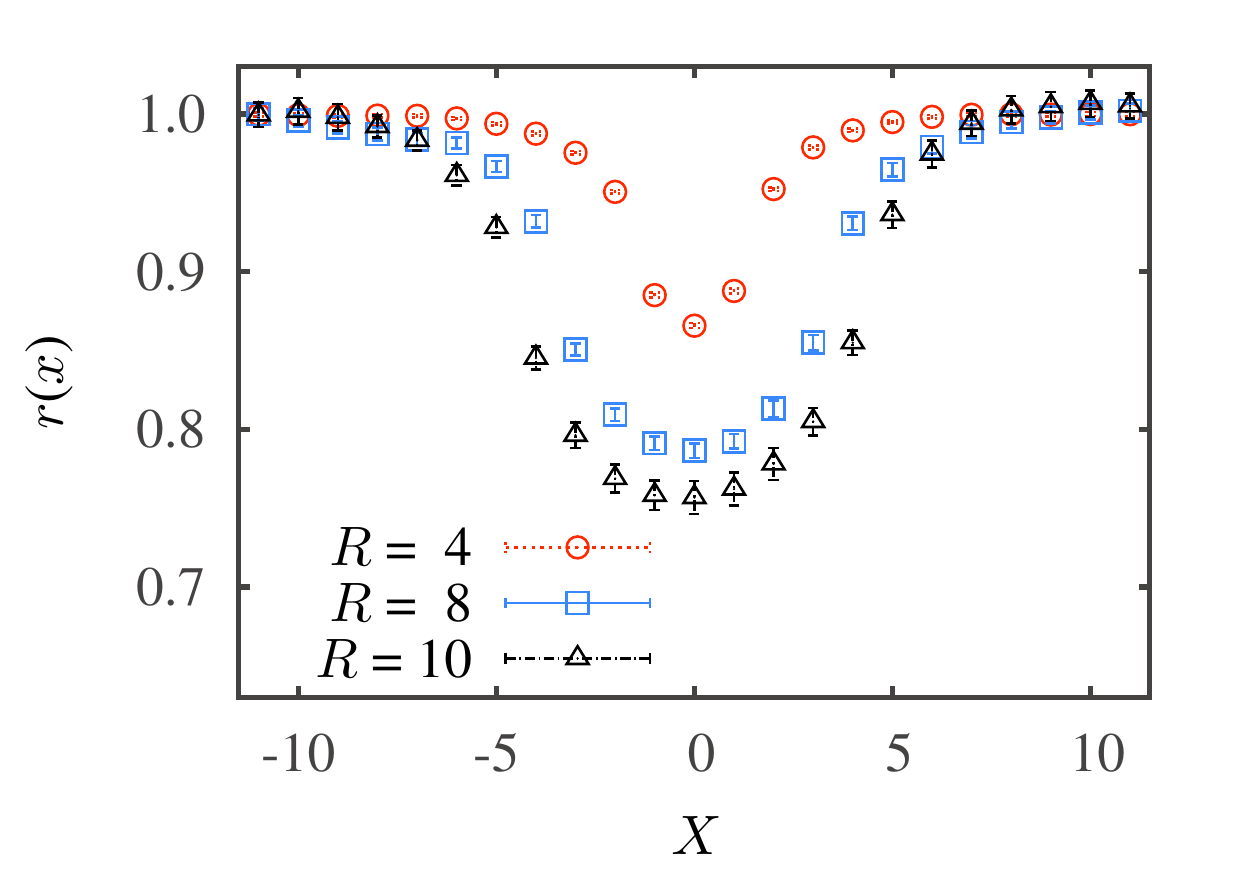}
  \includegraphics[width=0.49\textwidth,clip]{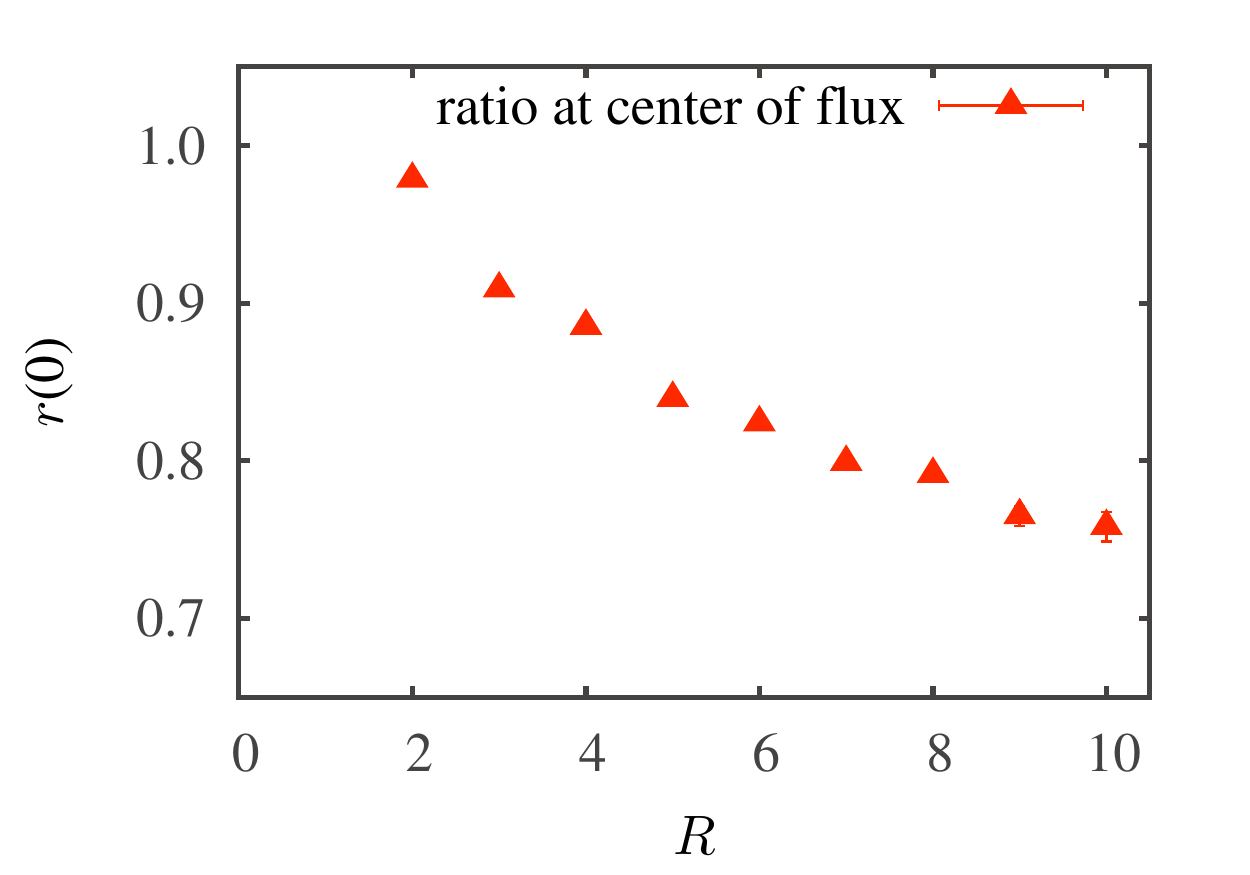}
  \caption{
    (a) The cross-section of chiral condensate ratio
    with color source separation at $R = 4, 8$, and $10$.
    (b) $R$-dependence of the ratio at the center of the flux.
    \label{fig:chiral_R_dependence}
  }
\end{figure}

We also show the $R$ dependence of the restoration.
Figure \ref{fig:chiral_R_dependence} (a)
is the cross-section of the ratio $r(\vec{x})$
along $X$-axis with $R = 4$, $8$ and $10$.
The positions of color sources are $X = \pm 2$, $\pm 4$, $\pm 5$, respectively.
It is clear that the magnitude of restoration becomes larger with the size of systems.
Fig.~\ref{fig:chiral_R_dependence} (b) shows 
the ratio at the center of sources $r(0)$, which decreases monotonically.
However, it is unlikely that chiral symmetry is completely restored inside the flux-tube,
since a flux-tube is broken through quark-antiquark pair creation.
Considering the string breaking scale of around $1$~fm \cite{Bali:2000gf},
the maximum reduction would be at most 30\%
in the $\mathrm{Q\bar{Q}}$-system.

\subsection{Chiral condensate in three quark system}

It is also possible to calculate the chiral condensate with multiple color sources.
Here, we consider three color sources corresponding to a \textit{baryon}.
Using the path-ordered product $U_k \equiv \prod_{\Gamma_k} e^{iag A_k}$
along the path $\Gamma_k$, the 3Q-Wilson loop is given by
$W_{\mathrm{3Q}} \equiv \frac{1}{3}\varepsilon_{abc} \varepsilon_{a'b'c'}
  U_1^{aa'} U_2^{bb'} U_3^{cc'}$,
which is composed to form a color singlet \cite{Takahashi:2001}.
The ratio of chiral condensate in the 3Q-system $r_\mathrm{3Q}(\vec{x})$
is given by substituting the 3Q-Wilson loop $W_\mathrm{3Q}$ for the Wilson loop $W(R,T)$ 
in Eq. (\ref{eq:ratioCondensate}) \cite{Iritani:2013}.

\begin{figure}
  \centering
  \includegraphics[width=0.49\textwidth,clip]{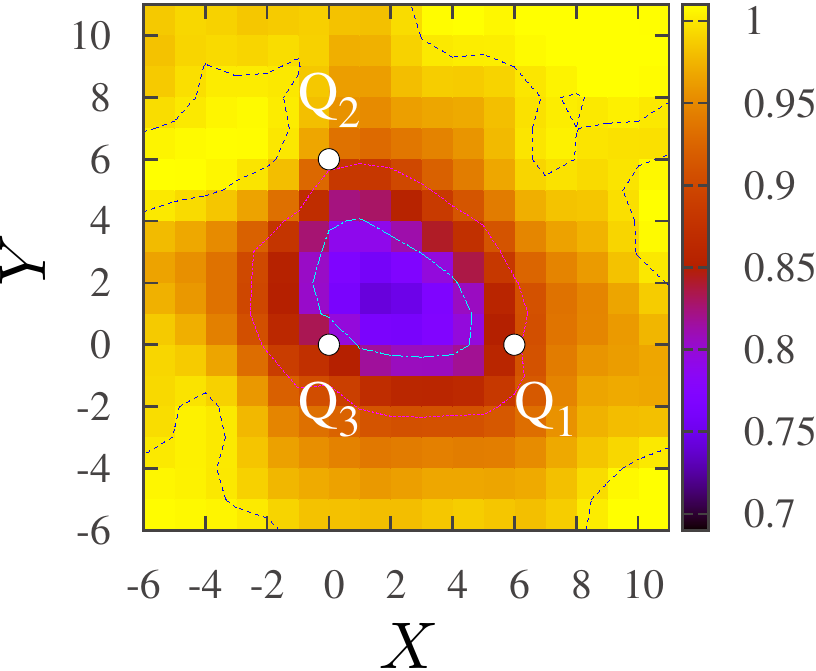}
  \includegraphics[width=0.49\textwidth,clip]{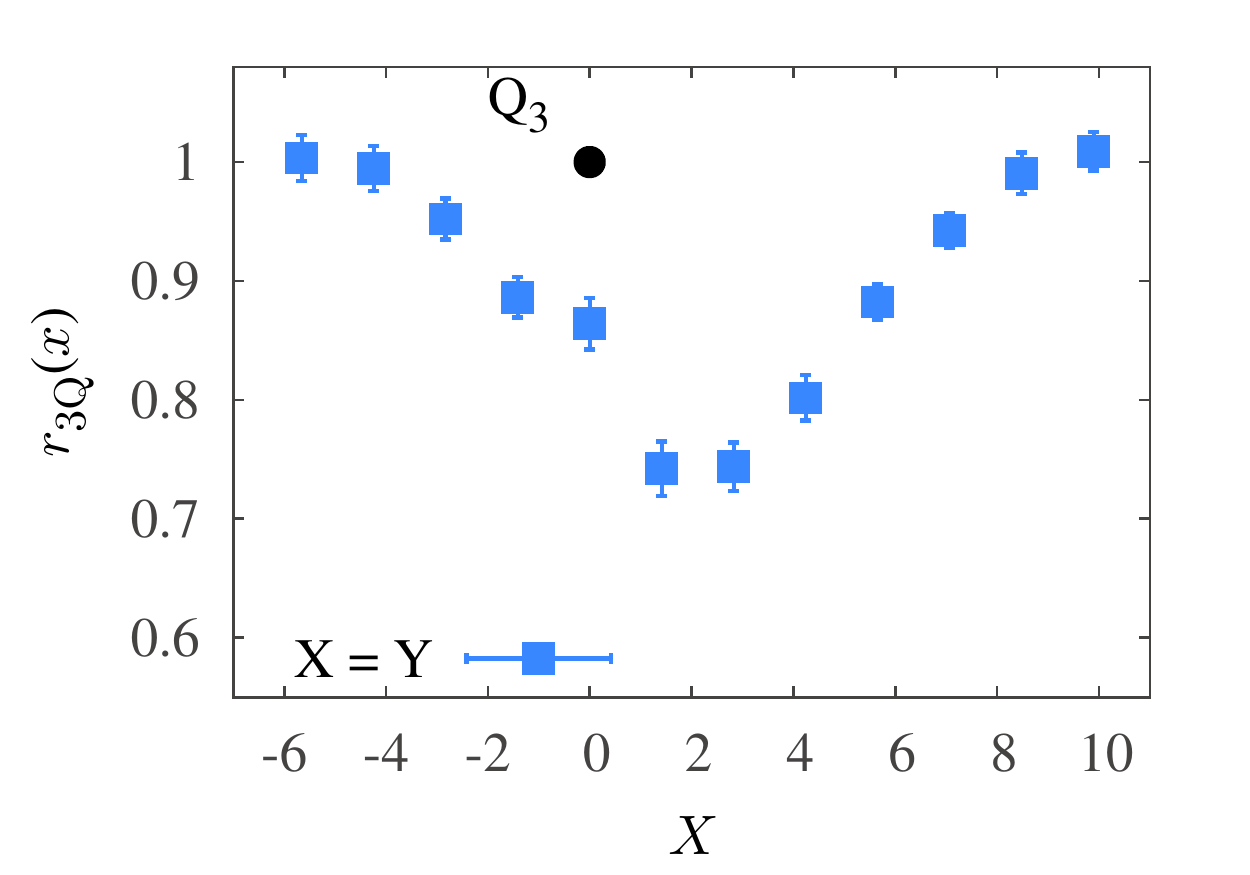}
  \caption{The ratio of chiral condensate around three-static color sources
    at $(X,Y) = (0,0)$, $(6,0)$ and $(0,6)$.
    (a) The heat-map of $r_\mathrm{3Q}(\vec{x})$.
    (b) The cross-section of $r_\mathrm{3Q}(\vec{x})$ along $X = Y$.
    \label{fig:chiral_3q_system}
  }
\end{figure}

We choose an isosceles right triangle configuration on $XY$-plane for simplicity.
Figure \ref{fig:chiral_3q_system} show the chiral condensate ratio 
$r_{\mathrm{3Q}}(\vec{x})$ and its cross-section along
$X = Y$ with color sources at $(X,Y) = (0,0)$, $(6,0)$ and $(0,6)$.
Similar to the $\mathrm{Q}\bar{\mathrm{Q}}$-system,
chiral symmetry is partially restored among color sources, 
where the magnitude of the condensate is reduced.
Around the center of sources,
the chiral condensate is reduced by about 30\%,
which is almost the same magnitude as the $\mathrm{Q}\bar{\mathrm{Q}}$ case
shown in Fig.~\ref{fig:chiral_ratio}.
We note that the characteristic $Y$-type flux is smeared 
in Fig.~\ref{fig:chiral_3q_system}, 
as the thickness of flux is comparable to the separation of color sources
\cite{Ichie:2002}.

\subsection{Partial restoration of chiral symmetry at ``finite-density''}
Finally, we discuss the partial restoration of chiral symmetry at finite density
\cite{Hayano:2008vn}.
Considering a single baryon in the spatial periodic box,
it could be regarded as a \textit{nuclear crystal}.
In this setup the baryon number density $\rho$ can be defined as $\rho \equiv 1/L^3$ 
with a spatial lattice size $L$.
Then, the \textit{nuclear matter density} is changed by the spatial volume of the box.
We calculate the net change of chiral condensate, which is given by the spatial average
of the ratio $r_\mathrm{3Q}(\vec{x})$ as
\begin{equation}
  \frac{\langle \bar{q}q \rangle_\rho}{\langle \bar{q}q \rangle_0}
  \equiv \frac{1}{L^3} \sum_{\vec{x}} 
  \frac{\langle \bar{q}q(\vec{x}) \rangle_\mathrm{3Q}}{\langle \bar{q}q \rangle}
  = \frac{1}{L^3} \sum_{\vec{x}} r_\mathrm{3Q}(\vec{x}).
  \label{eq:netChange}
\end{equation}
We use two lattice volumes $L^3 = 24^3$ and $16^3$.
These boxes correspond to $\rho \sim 0.3\rho_0$ and $\rho \sim \rho_0$, respectively,
with normal nuclear matter density $\rho_0 \simeq 0.18$~$\mathrm{fm}^{-3}$.
In this analysis, the size of a \textit{baryon} is determined by the shape of the 3Q-Wilson loop.
Here, we use the same spatial configuration as in Fig.~\ref{fig:chiral_3q_system}.

\begin{figure}
  \centering
  \includegraphics[width=0.59\textwidth,clip]{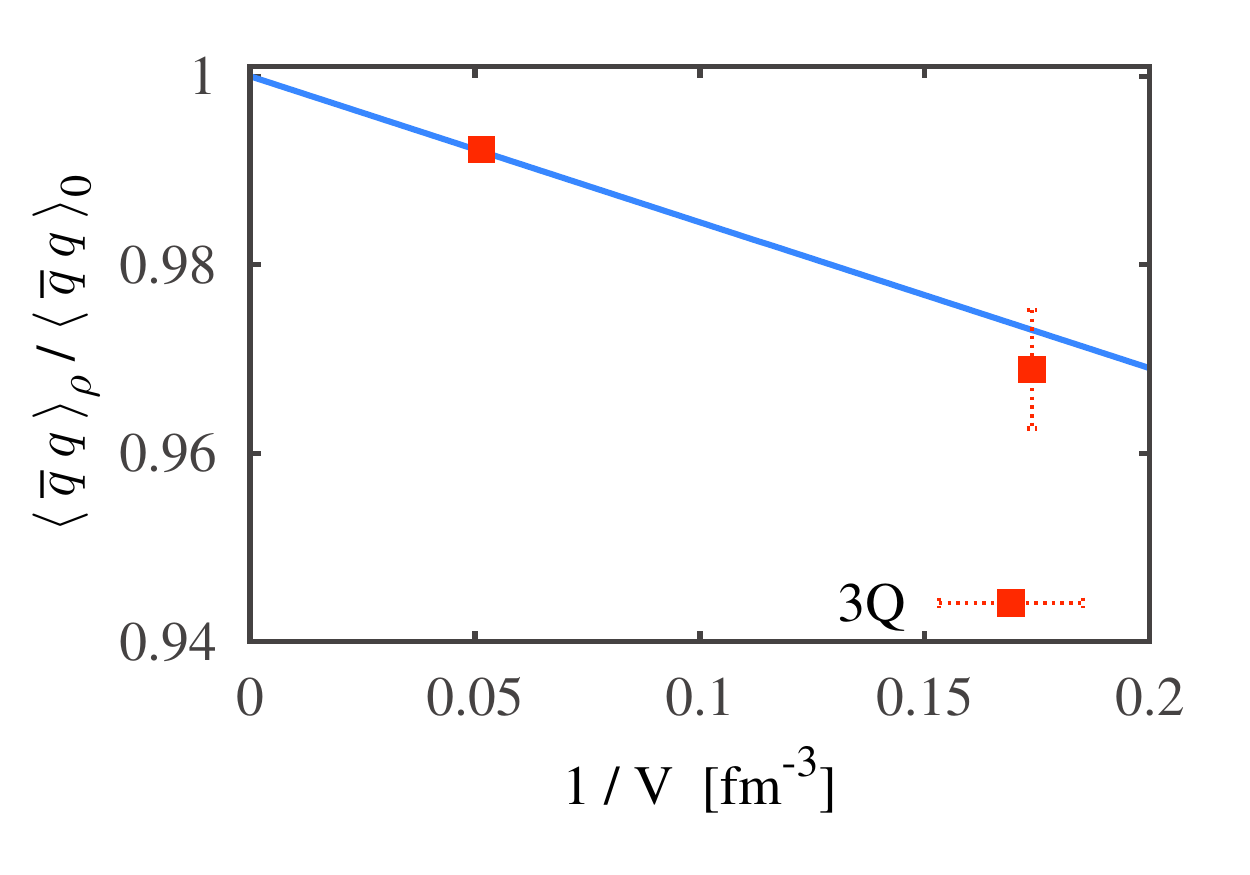}
  \caption{
    The net change of chiral condensate $\langle \bar{q}q \rangle_\rho/\langle \bar{q}q \rangle$
    with the spatial configuration of 3Q color sources 
    at $(X,Y) = (0,0)$, $(6,0)$, and $(0,6)$.
    We use the two spatial volume $24^3$ and $16^3$,
    which corresponds to $\rho \simeq 0.3\rho_0$ and $\rho_0$, respectively.
    \label{fig:chiral_finite_system}
  }
\end{figure}

Figure \ref{fig:chiral_finite_system} shows density dependence of the chiral condensate
$\langle \bar{q}q \rangle_\rho/\langle \bar{q}q \rangle_0$,
the solid line is a linear fit result from the vacuum expectation value.
In the setup shown in Fig.~\ref{fig:chiral_3q_system},
the restoration is estimated to be about 3\% at $\rho_0$.
Considering about 30\% of the chiral condensate
is expected to be reduced in the nuclear matter \cite{Hayano:2008vn},
our result seems small. 
However, the net change of the condensate in Eq.~(\ref{eq:netChange}) may depend 
on the size of the baryon.
For the setup of Fig.~\ref{fig:chiral_3q_system},
the root mean square radius is about 0.44~fm,
which is smaller than that of physical proton or 
neutron\footnote{For example, charge root mean square radius of proton is about 0.88~fm.}.
Similar to the $\mathrm{Q\bar{Q}}$-system shown in Fig.~\ref{fig:chiral_R_dependence},
the reduction becomes larger with the size of the 3Q-system.
Unfortunately, due to the large statistical errors,
it is difficult to numerically calculate 
larger 3Q-Wilson loop configurations than that in Fig.~\ref{fig:chiral_3q_system}.
The reduction of chiral condensate can be
more significant at a physical baryon size.

\section*{Acknowledgements}
  The lattice QCD calculations have been done on SR16000 at 
  High Energy Accelerator Research Organization (KEK)
  under a support of its Large Scale Simulation Program (No. 13/14-04).
  This work is supported in part by the Grant-in-Aid of the Japanese
  Ministry of Education (No. 26247043),
  and the SPIRE (Strategic Program for Innovative REsearch) Field5 project.


\begin{thebibliography}{99}


  \bibitem{Bali:2000gf} 
    G.~S.~Bali,
    \emph{Phys.\ Rept.}  {\bf 343} (2001) 1 [hep-ph/0001312].

  \bibitem{Bali:1995}
    G.~S.~Bali, K.~Schilling and C.~Schlichter,
    \emph{Phys.\ Rev.} {\bf D51} (1995) 5165 [hep-lat/9409005];
    P.~Cea and L.~Cosmai, 
      \emph{Phys.\ Rev.} {\bf D52} (1995) 5152 [hep-lat/9504008];
    R.~W.~Haymaker, V.~Singh, Y.~-C.~Peng and J.~Wosiek,
    \emph{Phys.\ Rev.} {\bf D53} (1996) 389 [hep-lat/9406021].

  \bibitem{Iritani:2013}
    T.~Iritani, G.~Cossu and S.~Hashimoto,
    \pos{PoS (Lattice 2013) 376} [arXiv:1311.0218 [hep-lat]];
    \pos{PoS (Hadron 2013) 159} [arXiv:1401.4293 [hep-lat]].

  \bibitem{Faber:1993}
    M.~Faber, M.~Schaler, and H.~Gausterer,
    \emph{Phys. Lett. B} {\bf 317} (1993) 409;
    W.~Sakuler, W.~Burger, M.~Faber, H.~Markum, M.~Muller, P.~De Forcrand,
    A.~Nakamura and I.~O.~Stamatescu,
    \emph{Phys.\ Lett.\ B} {\bf 276} (1992) 155;
    S.~Thurner, M.~Feurstein, H.~Markum, and W.~Sakuler,
    \emph{Phys.\ Rev.} {\bf D54} (1996) 3457;
    K.~H\"ubner, 
    \pos{PoS (Lattice 2007) 193} [arXiv:0709.1467 [hep-lat]].

  \bibitem{Suganuma:1990nn} 
    H.~Suganuma and T.~Tatsumi,
    \emph{Annals Phys.}\  {\bf 208} (1991) 470;
    \emph{Phys.\ Lett.\ B} {\bf 269} (1991) 371;
    \emph{Prog. Theor. Phys.} {\bf 90} (1993) 379.


  \bibitem{JLQCD}
    S.~Aoki et al. (JLQCD and TWQCD Collaborations), 
    \emph{Prog. Theor. Exp. Phys.} {\bf 2012} (2012) 01A106.

  \bibitem{GinspargWilson}
    P.~H.~Ginsparg and K.~G.~Wilson,
    \emph{Phys. Rev.} {\bf D25} (1982) 2649.

  \bibitem{Neuberger:1998}
    H.~Neuberger, \emph{Phys. Lett. B} {\bf 417} (1998) 141
    [hep-lat/9707022];
    \emph{ibid.} \textbf{427} (1998) 353 [hep-lat/9801031].

  \bibitem{BanksCasher}
    T.~Banks and A.~Casher, \emph{Nucl. Phys.} {\bf B169} (1980) 103.

  \bibitem{Gattringer:2002} 
    C.~Gattringer, 
    \emph{Phys. Rev. Lett.} {\bf 88} (2002) 221601 [hep-lat/020220].

  \bibitem{Schafer:1996wv}
    T.~Sch\"afer and E.~V.~Shuryak,
    \emph{Rev.\ Mod.\ Phys.} {\bf 70} (1998) 323 [hep-ph/9610451].

  \bibitem{Noaki:2009xi} 
    J.~Noaki, T.~W.~Chiu, H.~Fukaya, S.~Hashimoto, H.~Matsufuru, 
    T.~Onogi, E.~Shintani and N.~Yamada,
    \emph{Phys.\ Rev.} {\bf D81} (2010) 034502
    [arXiv:0907.2751 [hep-lat]].

  \bibitem{Takahashi:2001} 
    T.T.~Takahashi, H.~Matsufuru, Y.~Nemoto, and H.~Suganuma,
    \emph{Phys. Rev. Lett.} \textbf{86} (2001) 18;
    \emph{Phys. Rev.} \textbf{D65} (2002) 114509.

  \bibitem{Ichie:2002} 
    H.~Ichie, V.~Bornyakov, T.~Streuer and G.~Schierholz,
    \emph{Nucl.\ Phys.\ A} {\bf 7121} (2003) 899 [hep-lat/0212036];
    \emph{Nucl.\ Phys.\ Proc.\ Suppl.}  {\bf 119} (2003) 751 [hep-lat/0212024].

  \bibitem{Hayano:2008vn} 
    R.~S.~Hayano and T.~Hatsuda,
    \emph{Rev.\ Mod.\ Phys.}  {\bf 82} (2010) 2949 [arXiv:0812.1702 [nucl-ex]].

\end{thebibliography}
\end{document}